\documentclass{elsart}
\usepackage{amsmath}
\usepackage{graphics}

\newcommand{\beq}{ \begin{equation} }
\newcommand{\eeq}{ \end{equation} }
\newcommand{\non}{\nonumber}
\newcommand{\Dop}{\ensuremath{\mathcal{D}_{\text{op}}}}

\def\ket#1{|#1\rangle}
\def\bra#1{\langle#1|}
\def\tran#1#2{\langle#1|#2\rangle}

\def\mel#1#2#3{\langle#1|#2|#3\rangle}

\def\abs#1{\lvert#1\rvert}

\DeclareMathOperator{\asin}{asin}
\DeclareMathOperator{\acos}{acos}

\begin{document}

\begin{frontmatter}
\title{Tunneling Rate for Superparamagnetic Particles by the Instanton Method}
\author[UFMG]{E. H. Martins Ferreira},
%\ead{erlon@fisica.ufmg.br}
\author[UFMG,USP]{M. C. Nemes\thanksref{permad}}
%\ead{carolina@fisica.ufmg.br}
\address[UFMG]{Departamento de F\'\i sica, Universidade Federal de Minas Gerais,\\
CP 702, 30161-970 Belo Horizonte, MG, Brazil}
\address[USP]{Instituto de F\'{\i}sica, Universidade de S\~ao Paulo, \\ 
CP 66318, 05315-970 S\~ao Paulo, SP, Brazil}
\thanks[permad]{permanent address: \ref{UFMG}}

\begin{abstract}
We derive the tunneling rate for paramagnetic molecules in the context of a collective spin model. By means of path integral methods an analytical expression is derived. Given the very large spins in question ($s \sim 3000\hbar$), the observation of magnetization changes due to pure unitary tunnel effects is unlikely.
\end{abstract}

\end{frontmatter}

%-----------------------------------------------------------------------
\section{Introduction}
%-----------------------------------------------------------------------

The spin tunneling in magnetic materials has atracted much attention since experimental evidences
of this phenomenon were observed. Although tunneling is a purely quantum effect, it is interesting
to note how it still manifests in {\em superparamagnetic particles} of macroscopic size (of the order
of tenths of nanometers). These ``particles'' are generally magnetic grains or clusters, whose spins
are highly correlated, and they behave like they were a single large spin particle\cite{stamp,garg}.
The {\em Macroscopic Quantum Tunneling} (MQT) is the tunneling of a single macroscopic variable
(in the present case, the magnetization) through a barrier, between two minima of the effective
potential of a macroscopic system. For small magnetic clusters, these minima correspond to two
states of opposite magnetization (easy direction), and the barrier is proportional to the exchange
anisotropy. When there is a repeated coherent tunneling betwenn the two wells, then we have a case
of {\em Macroscopic Quantum Coherence} (MQC), and all spins have their relative orientation unchanged\cite{garcia}.

The existence of tunneling removes the degenerescence of the energy levels of the system, and the
tunneling rate is given by the energy difference between these levels. A useful and well known
technique for performing these calculations is the {\em instanton}\cite{coleman}, which is based
on the {\em Path Integral} formalism developed by Feynman\cite{feynman}. In fact, the application
of the method, together with the Villain transformation to the tunneling of spins has been introduced
by Enz and Schilling in ref \cite{enz} (see also the review by Schilling in \cite{schilling}).
This is a non-pertubative method that provides the energy of the fundamental states with great precision.
The proposal of this work is to show how to apply the instanton technique to the problem of the spin
tunneling of a superparamagnetic particle with a detailed discussion of all the necessary steps to find
the energy splitting of the lowest levels.

%-----------------------------------------------------------------------
\section{The Hamiltonian of the system in terms of canonical operators}
%-----------------------------------------------------------------------
We start here with a general Hamiltonian of a single large spin system with easy direction on the z-axis which has been used in the context of superparamagnetic particles\cite{pfannes}, given by
\beq
H = -AS_z^2 + B(S_+^2+S_-^2),
\end{equation}
with the anisotropy constants  $A$ and $B$ satisfying $A \gg B > 0$. $S_{z}$ and $S_{\pm} = S_{x} \pm iS_{y}$ are the standard spin operators. In the basis of the eigenvectors of $S^{2}$ and $S_{z}$, these spin operators satisfy
\begin{subequations}
\label{eq:autovetSz}
\begin{align}
S_{z}\ket{s,m} &= m\ket{s,m} \\
S_{\pm}\ket{s,m}&= \sqrt{(s\mp m)(s\pm m +1)}\ket{s,m \pm 1}
\end{align}
\end{subequations}

To solve this problem, we shall use the Villain transform\cite{villain}, which consists of defining the operators $S_+$ and $S_-$ in terms of $S_z$ and a new operator $\hat\varphi$,
\begin{subequations}
\begin{align}
S_{+} &= \e^{i\hat{\varphi}}\sqrt{s(s+1)-S_{z}(S_{z}+1)} \label{eq:sp} \\
S_{-} &= \sqrt{s(s+1)-S_{z}(S_{z}+1)}\,\e^{-i\hat{\varphi}}
\end{align}
\end{subequations}
with 
\begin{equation}
[\hat{\varphi},S_{z}] = i. \label{eq:comrel1}
\end{equation}

The eigenvectors of $\hat\varphi$, $\{\ket{\phi}\}$, are such that
\begin{equation} 
\tran{\phi}{m} = \frac{\e^{im\phi}}{\sqrt{2\pi}}. \label{eq:mphi}
\end{equation}
To expand $S_+$ and $S_-$ we need that $||S_z|| \ll s$. This would be true, were the system's easy direction not on the $z$-axis, but on the $y$ or $x$-axis. This direction is clearly arbitrary, so we choose it on the $y$-axis, and rewrite the Hamiltonian as 
\beq
H = -AS_{y}^{2} + 2B(S_{z}^{2}-S_{x}^{2}), \label{eq:hamsy} 
\end{equation}
or equivalently
\beq
%\begin{split}
H = \frac{A}{4}(S_{+}^{2}+S_{-}^{2}-S_{+}S_{-}-S_{-}S_{+}) + 2B(S_{z}^{2}-\frac{1}{4}(S_{+}^{2}+S_{-}^{2}+S_{+}S_{-}+S_{-}S_{+})) .
%\end{split}
\end{equation}
Besides the commutation relation \eqref{eq:comrel1}, we need another one, which can be derived from it and proves to be much useful, that is
\begin{equation}
[S_{z},\e^{\pm im\hat{\varphi}}] = \pm m\,\e^{\pm im\hat{\varphi}}. \label{eq:comrel2} 
\end{equation}
One implication of the above equation is
\begin{equation}
f(S_{z})\,\e^{\pm i\hat{\varphi}} = \e^{\pm i\hat{\varphi}}f(S_{z}\pm 1), \label{eq:fSz}
\end{equation}
where $f(S_{z})$ is any function of $S_{z}$. With these relations, we can write the Hamiltonian, in the ``normal order'', in terms of these new operators, that is
\begin{equation}
 \begin{split}
H & = A[(s(s+1)-S_{z}^{2})(\cos^{2}\hat{\varphi}-1)+iS_{z}\sin 2\hat{\varphi}] + \\
& \qquad 2B[S_{z}^{2}-(s(s+1)-S_{z}^{2})\cos^{2}\hat{\varphi} - iS_{z}\sin 2\hat{\varphi}].
\end{split}
\end{equation}

%-----------------------------------------------------------------------
\section{Transition amplitude}
%-----------------------------------------------------------------------

We have to compute the following matrix element
\begin{equation}
\tran{\phi_f,t_f}{\phi_i,t_i} =  \mel{\phi_f}{\e^{-iH(t_f-t_i)}}{\phi_i}.
\end{equation}
This is achieved in a standard fashion by discretizing the time in intervals $\epsilon = (t_f-t_i)/N$ and inserting different projection operators like$\int \d\phi_k \ket{\phi_k}\bra{\phi_k}$, between each time interval. We get
\beq
 \tran{\phi_{f},t_{f}}{\phi_{i},t_{i}} =  
\lim_{N \rightarrow \infty} \int\left(\prod_{k=1}^{N-1}\d\phi_{k}\right) \mel{\phi_{f}}{\e^{-i\epsilon H}}{\phi_{N-1}} \ldots \mel{\phi_{1}}{\e^{-i\epsilon H}}{\phi_{i}} \label{eq:phifphii}
\end{equation}

Each one of the inner products above is given by
\beq \label{eq:elemtrans} \begin{split}
\mel{\phi_{k}}{\e^{-i\epsilon H}}{\phi_{k-1}} & = \sum_{m_{k}=-s}^{s} \tran{\phi_{k}}{m_{k}} \mel{m_{k}}{\e^{-i\epsilon H}}{\phi_{k-1}}\\
& = \sum_{m_{k}=-s}^{s}\frac{\e^{im_{k}(\phi_{k}-\phi_{k-1})}}{2\pi}\ \e^{-i\epsilon H(m_{k},\phi_{k-1})}
\end{split} 
\end{equation}
with
\beq \begin{split}
H(m_{k},\phi_{k-1}) & \equiv \frac{\mel{\phi_{k}}{H}{\phi_{k-1}}}{\tran{\phi_{k}}{\phi_{k-1}}} = \\
& = A[(s(s+1)-m_{k}^{2})(\cos^{2}\phi_{k-1}-1)+im_{k}\sin 2\phi_{k-1}]\ + \\
& \qquad 2B[m_{k}^{2}-(s(s+1)-m_{k}^{2})\cos^{2}\phi_{k-1}-im_{k}\sin 2\phi_{k-1}]
\end{split}
\end{equation}
Grouping the $m_{k}$-terms and making the substitution
\beq 
m_{k} + i\frac{(A-2B)\sin 2\phi_{k-1}} {2[A+2B-(A-2B)\cos^{2}\phi_{k-1}]} \rightarrow p_{k} 
\end{equation}
we get, after transforming the summation into an integral and taking the limit $s \rightarrow \infty$ (see Eq. (2.13) of Johnson\cite{johnson}),
\beq 
\mel{\phi_{k}}{\e^{-i\epsilon H}}{\phi_{k-1}} = \int \frac{\d p_{k}}{2\pi} \e^{i(p_{k}+\Delta)(\phi_{k}-\phi_{k-1})}\ \e^{-i\epsilon H(p_{k},\phi_{k-1})} \label{eq:phikphik-1}
\end{equation}
where
\begin{align*}
H(p_{k},\phi_{k-1}) & \equiv \frac{p_{k}^{2}}{2M(\phi_{k-1})} + V(\phi_{k-1}) \\
M(\phi_{k-1}) & \equiv \frac{1}{2[A+2B-(A-2B)\cos^{2}\phi_{k-1}]} \\
V(\phi_{k-1}) & \equiv (A-2B)s(s+1)\cos^{2}\phi_{k-1}  - As(s+1)
\end{align*}
and $\Delta = 0$ if $s$ is integer or  $\Delta = \half $ if $s$ is half-integer.
Now we use Eq. \eqref{eq:phikphik-1} back into Eq. \eqref{eq:phifphii} and take the limit $N \rightarrow \infty$ to find
\beq \label{eq:intdphidp} 
\begin{split}
\langle \phi_{f},t_{f}|\phi_{i},t_{i}\rangle & = \lim_{N \rightarrow \infty} \int \left(\prod_{k=1}^{N-1}\d\phi_{k}\right) \left(\prod_{k=1}^{N}\frac{\d p_{k}}{2\pi}\right) \\
&\qquad \qquad \times \exp\left\{i\epsilon\sum_{k=1}^{N} (p_{k}+\Delta)\left(\frac{\phi_{k}-\phi_{k-1}}{\epsilon}\right)-H(p_{k},\phi_{k-1})\right\} \\
&= \int \mathcal{D}[\phi]\mathcal{D}[p] \exp\left\{i\int \d t\left[(p+\Delta)\dot{\phi}-\frac{p^{2}}{2M(\phi)}-V(\phi)\right]\right\}.
\end{split} \end{equation}

Completing square for $p_{k}$, and computing the gaussian integral, we get
\beq \label{eq:intdphi}
\langle \phi_{f},t_{f}|\phi_{i},t_{i}\rangle = \int \mathcal{D}[\phi(t)] \exp\left\{i\int \d t\left[\Delta\dot{\phi}+\frac{M(\phi)\dot{\phi}^{2}}{2}-V(\phi)\right]\right\}.
\end{equation}
Making a Wick rotation, we then have
\beq
\begin{split}
\mel{\phi_f}{\e^{-H(\tau_f-\tau_i)}}{\phi_i} & = \int \mathcal{D}[\phi(\tau)]\exp\left\{-\int \d\tau\left[-i\Delta\dot{\phi}+\frac{M(\phi)\dot{\phi}^{2}}{2}+V(\phi)\right]\right\} \\
& = \int \mathcal{D}[\phi(\tau)]\e^{-S_E}
\end{split} 
\end{equation}
from where we find the classical Hamiltonian and Lagrangian of the system
\begin{subequations}
\begin{align}
H(p,\phi) & = [(2B-A)\cos^{2}\phi+2B+A]p^{2}+(A-2B)s(s+1)\cos^{2}\phi - As(s+1) \\
L(\phi,\dot{\phi}) & = \frac{\dot{\phi}^{2}}{4[(2B-A)\cos^{2}\phi+2B+A]}-(A-2B)s(s+1)\cos^{2}\phi+As(s+1) 
\end{align}
\end{subequations}

%--------------------------------------------------------------
\section{Expansion of the action with variable mass}
%--------------------------------------------------------------

Let us take $s$ integer, so that $\Delta=0$, as it is appropriate for superparamagnetic particles. The case with $\Delta = \half$ leads to a topological selection rule, already discussed in the literature\cite{leuenberger}. The Euclidean action is then
\beq \label{eq:euclaction}
S_{E} = \int_{-T/2}^{T/2} \d\tau\ \tfrac{1}{2}M(\phi)\dot{\phi}^{2}+V(\phi). 
\end{equation}

The instanton is the solution which minimizes the Euclidean action. In the semi-classical limit, we make an expansion  of the action around the instanton $\tilde{\phi}$, taking
\beq
\phi = \tilde{\phi} + c_{i}\delta\varphi_{i}
\end{equation}
(repeated indices indicate summation). $\delta\varphi_{i}$ are eigenfunctions of an operator \Dop, and form an orthonormal basis with the boundary condition $\delta\varphi_{i}(\pm T/2) = 0$. Expanding each one of the terms above, we obtain
\begin{subequations}
\begin{align}
M(\phi) & = M(\tilde{\phi}+c_{i}\delta\varphi_{i}) = M(\tilde{\phi}) + \frac{\partial M}{\partial\phi}\Biggr\rvert_{\tilde{\phi}} c_{i}\delta\varphi_{i} + \frac{1}{2}\frac{\partial^{2}M}{\partial\phi^{2}}\Biggr\rvert_{\tilde{\phi}} c_{i}\delta\varphi_{i}c_{j}\delta\varphi_{j}\\
\dot{\phi}^{2}& = (\dot{\tilde{\phi}}+c_{i}\delta\dot{\varphi}_{i})^{2} = \dot{\tilde{\phi}}^{2} + 2\dot{\tilde{\phi}}c_{i}\delta\dot{\varphi}_{i} + c_{i}\delta\dot{\varphi}_{i}c_{j}\delta\dot{\varphi}_{j} \\
V(\phi)& = V(\tilde{\phi}) + \frac{\partial V}{\partial\phi}\Biggr\rvert_{\tilde{\phi}} c_{i}\delta\varphi_{i} + \frac{1}{2}\frac{\partial^{2}V}{\partial\phi^{2}}\Biggr\rvert_{\tilde{\phi}} c_{i}\delta\varphi_{i}c_{j}\delta\varphi_{j}
\end{align}
\end{subequations}
Substituting the expressions above into Eq. \eqref{eq:euclaction} and integrating by parts, the action becomes
\begin{align}
S[\phi] & = S[\tilde\phi] + \frac{1}{2}\int_{-T/2}^{T/2}\d\tau\sum_{i,j}c_{i}c_{j}\delta\varphi_{i} \left(-M(\tilde{\phi})\partial_{\tau}^{2} - \dot{\tilde{\phi}}\frac{\partial M}{\partial \phi}\partial_{\tau}-\frac{1}{2}\frac{\partial^{2}M}{\partial\phi^{2}}\dot{\tilde{\phi}}^{2} -
\ddot{\tilde{\phi}}\frac{\partial M}{\partial\phi} + \frac{\partial^{2}V}{\partial\phi^{2}}\right)\delta\varphi_{j} \non \\
&= S_0 + \frac{1}{2}\int_{-T/2}^{T/2}\d\tau \sum_{i,j}c_{i}c_{j}\delta\varphi_{i}\mathcal{D}_{\text{op}}\delta\varphi_{j}
\end{align}
The transition amplitude is then
\beq 
\mel{\phi_f}{\e^{-H(\tau_f-\tau_i}}{\phi_i} = \e^{-S_0}[\det(\Dop)]^{-1/2}.
\end{equation}

A chain of instantons and anti-instantons is also a solution which minimizes the action. When we take these solutions into account(see \cite{coleman}), we get finally
\beq 
\mel{\phi_f}{\e^{-HT}}{\phi_i} = \left(\frac{m\omega}{\pi}\right)^{1/2} \e^{-\omega T/2} \sum_{n\ \text odd} \frac{(K\e^{-S_{0}}T)^{n}}{n!}, \label{eq:continst} 
\end{equation}
where $T \equiv \tau_f-\tau_i$ and
\beq
K = \lim_{T\rightarrow\infty}\left(\frac{\det(\Dop)}{\det(-\partial_{\tau}^{2}+\omega^{2})}\right)^{-1/2}.
\end{equation}

%-----------------------------------------------------------------------
\section{The instanton}
\label{sec:instanton}
%-----------------------------------------------------------------------

To find the instanton, we write the Euclidean action as 
\beq S_E = \int \d\tau \bigl[-ip\dot\phi + H(p,\phi)\bigr]. \label{eq:classact} \end{equation}
Using simple variational calculus, we have that the $\delta p$ coeficient must vanish, and then
\beq
i\dot{\tilde{\phi}} = 2[(2B-A)\cos^{2}\tilde{\phi} +2B+A]\tilde{p}. \label{eq:iphipt}
\end{equation}
The condition that the instanton is the solution of minimal energy gives
\begin{align}
H(\tilde{p},\tilde{\phi}) &= -As(s+1) \non \\
\tilde{p}^{2} & = \frac{(2B-A)s(s+1)\cos^{2}\tilde{\phi}}{(2B-A)\cos^{2}\tilde{\phi} +2B+A}, \label{eq:p2} 
\end{align}
and so
\beq 
i\dot{\tilde{\phi}} = \pm 2\sqrt{2B-A}\sqrt{s(s+1)}|\cos\tilde{\phi}|\sqrt{(2B-A)\cos^{2}\tilde{\phi} +2B+A}
\end{equation}
We make $k \equiv \frac{2B-A}{2B+A}$, such that $-1<k<1$, and then 
\beq 
\dot{\tilde{\phi}} = \pm 2\sqrt{A^{2}-4B^{2}}\sqrt{s(s+1)}|\cos\tilde{\phi}|\sqrt{1+k\cos^{2}\tilde{\phi}}
\end{equation}
Now, integrating the equation above, we have
\beq \int \frac{\d\tilde{\phi}}{\cos\tilde{\phi}\sqrt{1+k\cos^{2}\tilde{\phi}}} = \pm \int 2\sqrt{s(s+1)}\sqrt{A^{2}-4B^{2}}\d\tau. \end{equation}
Making the substitution $x=\sin\tilde{\phi}$, $\d x=\cos\tilde{\phi}\,\d\tilde{\phi}$, we can readily solve the integral to find 
\beq \int \frac{\d x}{(1-x^{2})\sqrt{1+k(1-x^{2})}} = \frac{1}{2}\ln\left(\frac{\sqrt{1+k-kx^{2}}+x}{\sqrt{1+k-kx^{2}}-x}\right) \end{equation}

It is easy to check that $\omega \equiv \sqrt{V''(\phi_{0})/M(\phi_{0})}$ is equal to $2\sqrt{s(s+1)}\sqrt{A^{2}-4B^{2}}$ ($\phi_0$ is the initial value of $\phi(\tau)$ equal to $\phi(\pm \infty)=\pm \pi/2$), and therefore the solution is given by
\begin{align}
\tilde{\phi} & = \asin\left\{\frac{\sqrt{1+k}\sinh[\pm\omega(\tau-\tau_{0})]}{\sqrt{1+(1+k)\sinh^{2}[\pm\omega(\tau-\tau_{0})]}}\right\} \label{eq:solinstant} \\
\tilde{p}(\tau) & = \pm \sqrt{\frac{k}{1+k}}\frac{2\sqrt{s(s+1)}}{\cosh[\omega(\tau-\tau_{0})]} \label{eq:ptau}
\end{align}

\begin{figure}
\begin{center}
  \includegraphics{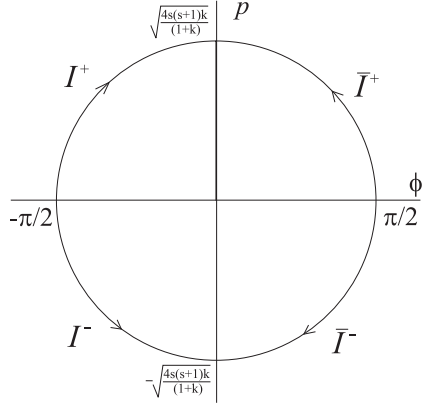}
\end{center}
\caption{All possible instanton and anti-instanton solutions}
\label{fig:instanti}
\end{figure}

There are thus two instanton-like solutions, with positive ($I^{+}$) and negative ($I^{-}$) signal of Eq. \eqref{eq:ptau}, and the corresponding anti-instantons ($\bar{I}^{+}$ and $\bar{I}^{-}$), given by the positive or negative signal of the argument of the function $\sinh$ in Eq. \eqref{eq:solinstant}. Figure \ref{fig:instanti} shows the different types of instantons and anti-instantons. They are necessary to the calculation of the Berry's phase later (section \ref{sec:faseberry}).

Making the identification $\tilde{p} = i\sqrt{s(s+1)}\cos(\tilde{\theta})$, we obtain the classical angle $\theta$ 
\beq \tilde{\theta}(\tau) = \acos\left\{\frac{2\sqrt{-k/(1+k)}}{\cosh[\omega(\tau-\tau_{0})]}\right\}. \end{equation}

\begin{figure}
\begin{center}
  \includegraphics{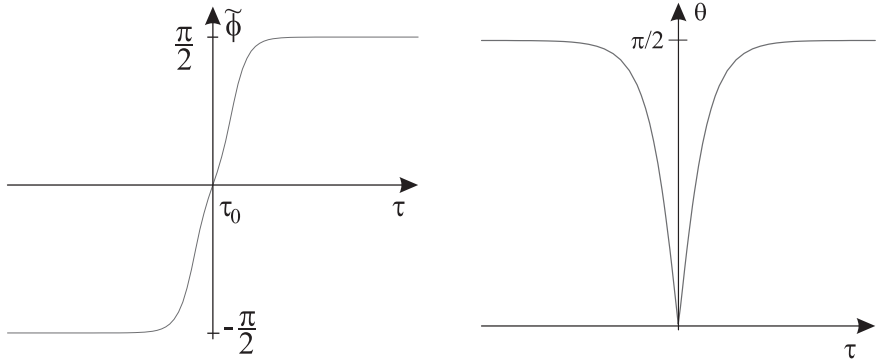}
\end{center}
\caption{Positive instanton solution for the spin tunneling}
\label{fig:solinstant}
\end{figure}

Figure \ref{fig:solinstant} shows the positive graphs of these functions. We can see clearly the expected result. For example, for the instanton represented in this picture ($I^{+}$), we have $\phi=-\pi/2$ and $\theta=\pi/2$ when $\tau\rightarrow-\infty$, and $\phi=\pi/2$ and $\theta=\pi/2$ when $\tau\rightarrow\infty$, which are the states of minimal energy. The classical action can now be readily computed
\begin{align}
S_{0} &= -i\int_{-T/2}^{T/2} \d\tau \dot{\tilde{\phi}}\,\tilde{p} 
 = -i\int_{-\pi/2}^{\pi/2} \d\tilde{\phi}\,\tilde{p} \non \\
S_{0}& = \sqrt{s(s+1)}\ln\left(\frac{1+\sqrt{-k}}{1-\sqrt{-k}}\right)
\end{align}

%--------------------------------------------------------------
\subsection{Eingenvalue zero}
%--------------------------------------------------------------

If we apply the operator \Dop\ to the function $\dot{\tilde{\phi}}$
\beq \begin{split}
\Dop\dot{\tilde{\phi}} & = -M(\tilde{\phi})\partial_{\tau}^{2}\dot{\tilde{\phi}} - \dot{\tilde{\phi}}\frac{\partial M}{\partial \phi}\partial_{\tau}\dot{\tilde{\phi}}-\frac{1}{2}\frac{\partial^{2}M}{\partial\phi^{2}}\dot{\tilde{\phi}}^{3} -
\ddot{\tilde{\phi}}\frac{\partial M}{\partial\phi}\dot{\tilde{\phi}} + \frac{\partial^{2}V}{\partial\phi^{2}}\dot{\tilde{\phi}} \\
& = \partial_{\tau}\left(-M(\tilde{\phi})\ddot{\tilde{\phi}} -\frac{1}{2}\dot{\tilde{\phi}}^{2}\frac{\partial M}{\partial\phi} + \frac{\partial V}{\partial\phi}\right)
\end{split}\end{equation}
we see the result is zero because the term in parantheses is simply the equation of motion of the system. Then we may conclude that $\dot{\tilde{\phi}}$ is the eigenfunction of \Dop\ with eigenvalue zero, and we must treat this eigenvalue separately. 

If $\tilde{\phi}$ centered at $\tau_{0}$ is an instanton solution, then $\tilde{\phi}$ centered at $\tau_{0}+a$ must also be. Thus
\beq  \tilde{\phi}_{\tau_{0}+a}(\tau) = \tilde{\phi}_{\tau_{0}}(\tau-a) \approx \tilde{\phi}_{\tau_{0}}(\tau) - a\dot{\tilde{\phi}} \equiv \tilde{\phi}_{\tau_{0}}(\tau) +c_{0}\delta\varphi_{0},
\end{equation}
with the normalization condition imposed by    
\beq \int_{-T/2}^{T/2}\d\tau\,\delta\varphi_{0}^{2} = N\int_{-T/2}^{T/2}\d\tau\,\dot{\tilde{\phi}}^{2} = 1 . \end{equation}
The integration over $c_{0}$ was already performed while computing $S_{0}$ and therefore must be taken off the determinant. However, we must correct the fact that the integration was over $\tau_{0}$, and consequently we need a correct normalization factor. A variation $\d\tau_{0}$ in the center of the instanton causes a variation 
\[ \Delta[\phi(\tau)] = \dot{\tilde{\phi}}\ \d\tau_{0}. \]
Since $\phi(\tau)=\tilde{\phi}(\tau) + c_{i}\delta\varphi_{i}$, a variation in $c_{0}$ associated to $\delta\varphi_{0}$, gives 
\[ \Delta[\phi(\tau)] = dc_{0}\delta\varphi_{0}. \]
As $\delta\varphi_{0} = 1/\sqrt{N}\dot{\tilde{\phi}}$, then $dc_{0} = \sqrt{N}\d\tau_{0}$, and
\beq K = \sqrt{\frac{N}{2\pi}} \lim_{T\rightarrow\infty}\left(\frac{\det'(\Dop)}{\det(-\partial_{\tau}^{2}+\omega^{2})}\right)^{-1/2}. \end{equation}
The prime indicates that the eigenvalue zero must not be computed in the determinant and the factor $1/\sqrt{2\pi}$ comes from the definition of $\mathcal{D}[\phi]=\prod_{n}dc_{n}/\sqrt{2\pi}$.

%--------------------------------------------------------------
\subsection{Calculation of K}
%--------------------------------------------------------------

From the theory of differential equations, the following relation\footnote{A prove of this relation may be found in Coleman\cite{coleman}, Appendix 1} can be derived  for the eigenvalue $\lambda=0$,
\beq \frac{\det(\Dop^{1})}{\det(\Dop^{2})} = \frac{\psi^{1}(T/2)}{\psi^{2}(T/2)}, \label{eq:detDop} \end{equation}
where $\Dop^{1,2}$ are  second order differential operators in the interval $[-T/2,T/2]$ and $\psi^{1,2}$ are eigenfunctions of these operatores, such that
\begin{align}
\psi^{1,2}(-T/2)=0 && \text{e} && \partial_{\tau}\psi^{1,2}(-T/2) = 1.
\label{eq:condpsi}
\end{align} 
We multiply and divide Eq. \eqref{eq:detDop} by $\lambda_{0}(T)$, which is the smallest eigenvalue of \Dop\ and vanishes in the limit $T\rightarrow\infty$. $\psi^{\omega}(\tau) = \frac{1}{\omega}\sinh[\omega(\tau+T/2)]$ is an eignefunction of the operator $-\partial_{\tau}^{2}+\omega^{2}$ which obeys the conditions above and has the form $\lim_{T\rightarrow\infty}\psi^{\omega}(T/2) = \e^{\omega T}/2\omega$ in the limit $T\rightarrow\infty$. So, in this limit, Eq. \eqref{eq:detDop} becomes
\beq
\frac{\det(\Dop)}{\det(-\partial_{\tau}^{2}+\omega^{2})}\frac{1}{\lambda_{0}(T)} = \frac{2\omega\psi(T/2)}{\lambda_{0}(T)\e^{\omega T}}.
\end{equation}

%--------------------------------------------------------------
\subsection{Calculation of $\psi(\tau)$}
%--------------------------------------------------------------

As already mentioned, $\dot{\tilde{\phi}}$ is an eigenfunction of the operator \Dop. Now we calculate this function explicitly starting from the instanton solution (Eq. \eqref{eq:solinstant}). Taking the derivative with respect to $\tau$, we see that
\[ \dot{\tilde{\phi}}(\tau) = \frac{\omega\sqrt{1+k}\cosh(\omega\tau)}{1+(1+k)\sinh^{2}(\omega\tau)}. \]
In the limit $\tau\rightarrow\pm\infty$, which is the limit of interest, this equation is simply
\[ \lim_{\tau\rightarrow\pm\infty} \dot{\tilde{\phi}}(\tau) = \frac{2\omega}{\sqrt{1+k}}\e^{-\omega\abs{\tau}} \]
and after normalization, we have
\beq 
\dot{\tilde{\phi}}_{N}(\tau) = \frac{1}{\sqrt{N}}\frac{2\omega}{\sqrt{1+k}}\e^{-\omega\abs{\tau}} \equiv C\e^{-\omega\abs{\tau}}=x_{1}. 
\end{equation}
This function however doesn't fulfill the conditions imposed by Eq. \eqref{eq:condpsi}, because $\partial_{\tau}\dot{\tilde{\phi}}_{N}(\tau=-\infty) = 0$. We then seek a solution $y_{1}$, linearly independent of $x_{1}$, such that $\psi = My_{1} + Nx_{1}$ satisfy the boundary conditions. We write $\Dop = \partial^{2} + I(\tau)\partial + J'(\tau)$. If $x_{1}$ and $y_{1}$ are eigenfunctions of \Dop\ with eigenvalue zero, so
\beq
\begin{cases}
y_{1}[\partial^{2}x_{1}+I(\tau)\partial x_{1} + J'(\tau)x_{1}]& = 0 \\
x_{1}[\partial^{2}y_{1}+I(\tau)\partial y_{1} + J'(\tau)y_{1}]& = 0 \\
\end{cases}
\end{equation}
Subtracting both equations
\[
\partial(y_{1}\partial x_{1}-x_{1}\partial y_{1})+I(\tau)[y_{1}\partial x_{1}-x_{1}\partial y_{1}] = 0 
\]
and definig $W(x_{1},y_{1}) = y_{1}\partial x_{1}-x_{1}\partial y_{1}$, the Wronskian of the system and $\partial F/F = I(\tau)$, we see that 
\beq \begin{split}
\partial W(x_{1},y_{1}) + \frac{\partial F}{F}W(x_{1},y_{1}) &= 0 \\
\partial(FW)& = 0 \Rightarrow FW = \text{constant}.
\end{split} \end{equation}
As $\Dop = -M\partial_{\tau}^{2} - (\partial_{\tau}M)\partial_{\tau} + J(\tau)$, we can identify $F$ with $M(\tilde{\phi})$, and then
\[ M(\tilde{\phi})W(x_{1},y_{1}) = \text{constant}. \]
We normalize $W(x_{1},y_{1})$ such that $M(\tilde{\phi})W(x_{1},y_{1})=2M_{0}\omega C^{2}$, where $M_{0}$ is the limiting value of $M(\phi)$ when $\tau \rightarrow \pm \infty$. Thus, in this limit, $W(x_{1},y_{1})=2\omega C^{2}$. Using the Wronskian, we get,
in the limit $\tau \rightarrow \pm \infty$,
\[ y_{1} = \pm C\e^{\pm\omega\tau}, \]
or equivalently
\beq y_{1} = \frac{\tau}{\abs{\tau}}C\e^{\omega\abs{\tau}}.
\end{equation}
We need now to find the coeficients $M$ and $N$, in order that $\psi$ satisfy the boundary conditions. They are
\[
M = \frac{\e^{-\omega T/2}}{2\omega C} \quad \text{and} \quad N = \frac{\e^{\omega T/2}}{2\omega C}.
\]
From the above, we finally have
\beq \psi(\tau) = \frac{1}{2\omega}\left[\e^{\omega T/2}\e^{-\omega\abs{\tau}}+\frac{\tau}{\abs{\tau}}\e^{-\omega T/2}\e^{\omega\abs{\tau}} \right] \end{equation}
and therefore we find the value of $\psi(T/2) = 1/\omega$.

%--------------------------------------------------------------
\subsection{Calculation of $\lambda_{0}$}
%--------------------------------------------------------------

We start with the eigenvalue equation of the operator \Dop,
\[ [-M\partial_{\tau}^{2} - (\partial_{\tau}M)\partial_{\tau} + J(\tau)]\psi_{\lambda}(\tau) = \lambda\psi_{\lambda}(\tau) \]
We write $\psi_{\lambda}(\tau)$ in an integral form like
\beq \psi_{\lambda}(\tau) = \psi(\tau) - D\int_{-T/2}^{\tau}\d s[y_{1}(\tau)x_{1}(s)-x_{1}(\tau)y_{1}(s)]\psi_{\lambda}(s) \label{eq:psilambda} \end{equation}
where $D$ is such that the solution above satisfy the eigenvalue equation. Taking the first and second order derivatives of $\psi_{\lambda}(\tau)$, we have
\[ \begin{split}
\psi_{\lambda}'(\tau)& = \psi'(\tau) - D\int_{-T/2}^{\tau}\d s[y_{1}'(\tau)x_{1}(s)-x_{1}'(\tau)y_{1}(s)]\psi_{\lambda}(s) \\
& \quad -D\underbrace{[y_{1}(\tau)x_{1}(\tau)-x_{1}(\tau)y_{1}(\tau)]}_{=0}\psi_{\lambda}(\tau) \\
\psi_{\lambda}''(\tau)& = \psi''(\tau) - D\int_{-T/2}^{\tau}\d s[y_{1}''(\tau)x_{1}(s)-x_{1}''(\tau)y_{1}(s)]\psi_{\lambda}(s) \\
& \quad -D\underbrace{[y_{1}'(\tau)x_{1}(\tau)-x_{1}'(\tau)y_{1}(\tau)]}_{=W(x_{1},y_{1})}\psi_{\lambda}(\tau).
\end{split} \]
Now, we substite it back into the eigenvalue equation, and use the equation of
motion, $-M\psi'(\tau) - (\partial_{\tau}M)\psi'(\tau) + J(\tau)\psi(\tau)=0$,
to get
\begin{multline}
[\lambda-MDW(x_{1},y_{1})]\psi_{\lambda}(\tau)= D\int_{-T/2}^{\tau}\d s\Bigl\{M[y_{1}''(\tau)x_{1}(s)-x_{1}''(\tau)y_{1}(s)]\\
+(\partial_{\tau}M)[y_{1}'(\tau)x_{1}(s)-x_{1}'(\tau)y_{1}(s)]-J(\tau)[y_{1}(\tau)x_{1}(s)-x_{1}(\tau)y_{1}(s)]\Bigr\}\psi_{\lambda}(s)
\end{multline}
The term on right is equivalent to
\beq
D\int_{-T/2}^{\tau}\d s\Bigl\{[-M\partial_{\tau}^{2} - (\partial_{\tau}M)\partial_{\tau} + J(\tau)]
[y_{1}(\tau)x_{1}(s)-x_{1}(\tau)y_{1}(s)]\Bigl\}\psi_{\lambda}(s)
\end{equation}
As $x_{1}(\tau)$ and $y_{1}(\tau)$ are eigenfunctions of \Dop\ with eigenvalue
zero, then the right side is null. Thus \begin{gather}
[\lambda-MDW(x_{1},y_{1})]\psi_{\lambda}(\tau) = 0 \nonumber \\ D =
\frac{\lambda}{MW(x_{1},y_{1})} \xrightarrow{T\rightarrow\infty}
\frac{\lambda}{2M_{0}\omega C^{2}} \end{gather}

Iterating Eq. \eqref{eq:psilambda}, we have
\beq \psi_{\lambda}(\tau) = \psi(\tau) - D\int \d s[\dotsb]\psi(s) + D^{2}\int \d s[\dotsb]\int \d s'[\dotsb]\psi(s') + \dotsb \end{equation}
Since $\lambda\rightarrow 0$, then $D^{2}\ll D$, and thus
\[
\psi_{\lambda}(\tau) \approx \psi_{0}(\tau) -
\frac{\lambda}{2M_{0}\omega C^{2}}\int_{-T/2}^{\tau} \d s[y_{1}(\tau)x_{1}(s)-x_{1}(\tau)y_{1}(s)]\psi_{0}(s).
\]
Using the results for $x_{1}$ and $y_{1}$, we have, for $\tau = T/2$
\beq
\psi_{\lambda}(T/2) = \frac{1}{\omega} - \frac{\lambda}{4M_{0}\omega^{2}C^{2}}\int_{-T/2}^{T/2} \d s[\e^{\omega T}x_{1}^{2}(s)-\e^{-\omega T}y_{1}^{2}(s)].
\end{equation}
This integral can now be readly solved and gives a result equal to $C^{2}/\omega \e^{\omega T}$. Therefore
\begin{gather}
\frac{1}{\omega} - \frac{\lambda_0 C^{2}/\omega \e^{\omega T}}{4M_{0}\omega^{2}C^{2}} = 0 \nonumber \\
\lambda_{0} = 4M_{0}\omega^{2}\e^{-\omega T}
\end{gather}

%--------------------------------------------------------------
\subsection{Value of $K$}
%--------------------------------------------------------------

With the results obtained here so far, the value of $K$ may be readily computed
\begin{align*}
K &= \sqrt{\frac{N}{2\pi}} \lim_{T\rightarrow\infty}\left(\frac{\det'(\Dop)}{\det(-\partial_{\tau}^{2}+\omega^{2})}\right)^{-1/2} =
 \sqrt{\frac{N}{2\pi}} \lim_{T\rightarrow\infty}\left(\frac{2\omega\psi(T/2)}{\lambda_{0}(T)\e^{\omega T}}\right)^{-1/2} = \\
&= \sqrt{\frac{N}{2\pi}} \sqrt{2M_{0}\omega^{2}}.
\end{align*}
$M_{0}$ is given by $\lim_{\phi\rightarrow \pm\pi/2}M(\phi)= [2(A+2B)]^{-1}$. $N$ is given by the normalization condition of $x_{1}$, and may be solved to give $N = 4\omega/(1+k)$. $1+k= 4B/(2B+A)$ and $\omega = 2\sqrt{s(s+1)}\sqrt{A^{2}-4B^{2}}$. Hence the final result for $K$ is
\beq K = \frac{\omega^{3/2}}{\sqrt{2\pi B}}. \end{equation}

%--------------------------------------------------------------
\section{Berry's phase} 
\label{sec:faseberry}
%--------------------------------------------------------------

As seen in section \ref{sec:instanton}, the model has two instanton-like solutions and their two corresponding anti-instanton. Then, there are  two possible paths to go from one minimum state to the other and two others to return. The action along any of the paths is the same up to a geometrical phase (Berry's phase) which is now computed. This phase is given by 
\[ \Phi_{B} = \exp \left\{\int \d\tau\, is\dot{\phi}\right\} = \exp \left\{\int \d\phi\,is\right\} \]
For the instanton $I^{+}$ and the anti-instanton $\bar{I}^{-}$, the Berry phase is
\begin{subequations}
\beq \Phi_{B}(I^{+}) = \Phi_{B}(\bar{I}^{-}) = \exp \left\{\int_{-\pi/2}^{\pi/2} \d\phi\,is\right\} = \e^{i\pi s} \label{eq:phiB1} \end{equation}
and to the instanton $I^{-}$ and the anti-instanton $\bar{I}^{+}$, 
\beq \Phi_{B}(I^{-}) = \Phi_{B}(\bar{I}^{+}) = \exp \left\{\int_{\pi/2}^{-\pi/2} \d\phi\,is\right\} = \e^{-i\pi s} \label{eq:phiB2} \end{equation}
\end{subequations}

We must introduce these phases in the calculation of Eq. \eqref{eq:continst}, because the contribution of each instanton is now multiplied by this phase. Taking account of the four possibilities of a 2-link chain (back and forth), the summation in Eq. \eqref{eq:continst} becomes
\begin{multline}
\frac{(K\e^{-S_{0}}\Phi_{B}(I^{+})T)(K\e^{-S_{0}}\Phi_{B}(\bar{I}^{+})T)}{2} + \frac{(K\e^{-S_{0}}\Phi_{B}(I^{-})T)(K\e^{-S_{0}}\Phi_{B}(\bar{I}^{-})T)}{2} \\
+\frac{(K\e^{-S_{0}}\Phi_{B}(I^{+})T)(K\e^{-S_{0}}\Phi_{B}(\bar{I}^{-})T)}{2}+\frac{(K\e^{-S_{0}}\Phi_{B}(I^{-})T)(K\e^{-S_{0}}\Phi_{B}(\bar{I}^{+})T)}{2}\\
= \frac{(K\e^{-S_{0}}T)^{2}(\Phi_{B}(I^{+}) + \Phi_{B}(I^{-}))(\Phi_{B}(\bar{I}^{+}) + \Phi_{B}(\bar{I}^{-}))}{2}
\end{multline}

With the results obtianed in \eqref{eq:phiB1} and \eqref{eq:phiB2}, we see that
\beq (\Phi_{B}(I^{+}) + \Phi_{B}(I^{-}))(\Phi_{B}(\bar{I}^{+}) + \Phi_{B}(\bar{I}^{-})) = 4\cos^{2}(\pi s) \end{equation}
and therefore in Eq. \eqref{eq:continst}, we must replace $\e^{-S_{0}/\hbar}$ by $2\cos(\pi s)\e^{-S_{0}/\hbar}$

%--------------------------------------------------------------
\section{Splitting of the ground state}
%--------------------------------------------------------------

Now we can finally calculate the energy splitting of the ground state due to the ressonant tunelling between them. Summing up Eq. \eqref{eq:continst}, we have
\beq 
\mel{\phi_f}{\e^{-HT}}{\phi_i} = \left(\frac{m\omega}{\pi}\right)^{1/2} \e^{-\omega T/2} \frac{1}{2}\left(\e^{2K\e^{-S_0}T\cos\pi s}+ \e^{-2K\e^{-S_0}T\cos\pi s}\right)
\end{equation}
from what we get the energies of the two lowest states to be
\[ E_\pm = \frac{1}{2}\omega \pm 2K\e^{S_0}\cos \pi s \]
Then the splitting in energy is
\beq \begin{split}
\Delta E_{\text{inst}} &= E_{+}-E_{-} = 4K\e^{-S_{0}}\cos \pi s \\
\Delta E_{\text{inst}} &= 4\left(\frac{\omega}{2\pi B}\right)^{1/2}\left(\frac{1+\sqrt{-k}}{1-\sqrt{-k}}\right)^{-\sqrt{s(s+1)}}\cos \pi s
\end{split} \end{equation}
Replacing the values of $\omega$ and $k$ in terms of $A$, $B$ and $s$, and using that $s \gg 1$, such that $\sqrt{s(s+1)} \approx s + \frac{1}{2}$, the final result will be
\beq
\Delta E_{\text{inst}} = \frac{8}{\sqrt{\pi B}}[s(s+1)(A^{2}-4B^{2})]^{3/4}\left(\frac{\sqrt{A+2B}+\sqrt{A-2B}}{\sqrt{A+2B}-\sqrt{A-2B}}\right)^{-(s+1/2)}\cos \pi s
\end{equation}

If we take the limit $B/A \ll 1$ and $ s \gg 1$, the equation above is equal to
\beq \Delta E_{\text{inst}} = \frac{8As^{3/2}}{\sqrt{\pi}}\left(\frac{B}{A}\right)^{s}\cos \pi s . \label{eq:deltaE} \end{equation}

This result may be compared with the expression obtained by Hartmann-Boutron\cite{boutron} using other method and with the rigorous universal upper bound for the tunneling rate of large quamtum spins recently obtained\cite{scharf} and independent of the form of the anisotropy, verifying an excellent agreement with both results.

We can still test the result, comparing it with the exact value obtained via diagonalization of the Hamiltonian. In Table \ref{tab:resultnum} we contrast these results, and it can be seen that, although the analytical calculation is valid for $s \gg 1$, even for small $s$, the discrepance between the values is small. 

%----------------------------------------------------------------------
\section{Conclusion}
%----------------------------------------------------------------------

We have found an analytical expression for the tunneling rate of a superparamagnetic particle using the instanton method and  compared it with others results obtained by different methods, and also with numerical calculations. The result proves to be in good agreement with all of them. It is worth noticing that for a typical superparamagnetic particle, with $s \sim 3000$ and $f \sim 10^{-2}$, the tunneling rate is practically zero, and then, no tunneling should be observed in practice. However, the change in magnetization direction of such particles is observed, what suggests that we must take the interaction with the environment into account. Work along these lines is in progress.

\begin{ack}
This work is supported by Brazilian research agencies CNPq and FAPESP.
\end{ack}

%\bibliographystyle{elsart-num}
%\bibliography{Artigo-PhysA}

\newpage

\begin{table}
\caption[C\'alculo de $\Delta E_{0}$]{}
{Comparison between the numerical and calculated results for the ground state energy splitting for different values of $s$. Here we took $f=0,2$. \smallskip}
\begin{center}
\begin{tabular}{ccc}
\hline \hline
s & $\Delta E_{0}^{\text{calc}}/A$ & $\Delta E_{0}^{\text{num}}/A$ \\
\hline
 5 & $5,61125\times 10^{-4}$  & $5,02226\times 10^{-4}$  \\
 8 & $1,08198\times 10^{-6}$  & $1,06013\times 10^{-6}$  \\
10 & $1,48684\times 10^{-8}$  & $1,51767\times 10^{-8}$  \\
12 & $1,93224\times 10^{-10}$ & $2,04068\times 10^{-10}$ \\
\hline \hline
\end{tabular}
\end{center}
\label{tab:resultnum}
\end{table}

\end{document}